\documentclass[a4paper,fleqn]{mnras}

\usepackage[T1]{fontenc}
\usepackage{ae,aecompl}

\usepackage{graphicx}	
\usepackage{amsmath}	
\usepackage{amssymb}	

\newcommand{\beq}{\begin{equation}}
\newcommand{\eeq}{\end{equation}}

\title[Radiation driven, optically thick winds]{The dynamics of radiation driven, optically thick winds}

\author[Shen, Nakar \& Piran]{
Rong-Feng Shen,$^{1,2}$\thanks{E-mail: shenrf3@mail.sysu.edu.cn}\thanks{Present address: Institute of Astronomy and Space Science, Sun Yat-Sen University, Guangzhou 510275, P. R. China}
Ehud Nakar,$^{2}$
and Tsvi Piran$^{1}$
\\
$^{1}$Racah Institute of Physics, Hebrew University of Jerusalem, Jerusalem 91904, Israel\\
$^{2}$Department of Astrophysics, School of Physics and Astronomy, Tel Aviv University, Tel Aviv 69978, Israel
}

\date{Accepted XXX. Received YYY; in original form ZZZ}

\pubyear{2016}

\begin{document}
\label{firstpage}
\pagerange{\pageref{firstpage}--\pageref{lastpage}}
\maketitle

\begin{abstract}
Recent observation of some luminous transient sources with low color temperatures suggests that the emission is dominated by optically thick winds driven by super-Eddington accretion. We present a general analytical theory of the dynamics of radiation pressure-driven, optically thick winds. Unlike the classical adiabatic stellar wind solution whose dynamics are solely determined by the sonic radius, here the loss of the radiation pressure  due to photon diffusion also plays an important role. We identify two high mass loss rate regimes ($\dot{M} > L_{\rm Edd}/c^2$). In the large total luminosity regime the solution resembles an adiabatic wind solution. Both the radiative luminosity, $L$,  and the kinetic luminosity, $L_k$,  are super-Eddington with  $L < L_k$ and $L \propto L_k^{1/3}$. In the lower total luminosity regime most of the energy is carried out by the radiation with $L_k < L \approx L_{\rm Edd}$.  In a third, low mass loss regime ($\dot{M} < L_{\rm Edd}/c^2$), the wind becomes optically thin early on and, unless gas pressure is important at this stage, the solution is very different from the adiabatic one. The results are independent from the energy generation mechanism at the foot of the wind, therefore they are applicable to a wide range of  mass ejection systems, from black hole accretion, to planetary nebulae, and to classical novae.
\end{abstract}




\section{Introduction}

Radiation-driven mass ejection (in the form of outflows, winds or shells) exists or was postulated to exist in different astronomical systems with different sources of energy generation. It was expected to take place in super-Eddington accreting black holes (e.g., Blandford \& Begelman 2004; Begelman 2012) as was confirmed numerically recently (e.g., Ohsuga et al. 2005; Yang et al. 2014). It was suggested to be the mechanism responsible for ejecting planetary nebula shells (Faulkner 1970; Finzi \& Wolf 1971), where the energy source is the stellar radiation during the star's AGB phase. Mass ejection is also responsible for the bright outbursts of classical novae (e.g., Ruggles \& Bath 1979) whose energy source is the runaway burning at the surface of a white dwarf. For all those systems, the ejected mass is dense enough so that it is optically thick to the photons generated at the center. Thus the ejected mass blocks the emission from the central source. The observed emission is generated, instead, by reprocessing the radiation from within.  

This wind-dominated emission scenario has been recently called for in multiple astrophysical contexts. In particular, there is a long standing debate (e.g., Fabbiano 2005) concerning the nature of the ultra-luminous X-rays sources (ULXs): X-ray sources with luminosity $L > 10^{39}$ erg s$^{-1}$. Are those ULXs stellar mass black holes (Poutanen et al. 2007; Gladstone, Roberts \& Done 2009), or are they the elusive intermediate-mass black holes (IMBHs, 10$^2$ -- 10$^4 M_{\odot}$) (Madau \& Rees 2001; Portegies Zwart et al. 2004; Miller et al. 2004, Kong et al. 2004; Miniutti et al. 2006; Liu \& Di Stefano 2008)? The new kinetics measurement of the companion of the ULX source X-1 in galaxy M101 constrains the black hole to be of a few tens of solar mass unless the viewing angle is unreasonably small (Liu et al. 2013). However, the low color temperature ($kT \simeq$ 0.1 keV) of this source suggests that the emission radius is at least 100 times larger than the innermost stable circular orbit of such a black hole. This source is a representative of a subclass of ULXs, referred to as ultra-luminous super-soft sources (Urquhart \& Soria 2016). We have recently shown (Shen et al. 2015; also supported by Soria \& Kong 2015) that an optically thick wind emerging from an accretion disk around a $\sim 10 M_{\odot}$ black hole can resolve this paradox. For an alternative model, see Gu et al. (2016).

Stellar tidal disruption events (TDEs) by supermassive black holes are super-Eddington accretion transients. TDE candidates detected in optical / UV wavelengths show color temperatures ($\sim 10^4$ K) that are generally lower by at least one order of magnitude than predicted by the standard accretion theory of a bare disk (van Velzen et al. 2011; Gezari et al. 2012; Arcavi et al. 2014; Chornock et al. 2014; Holoien et al. 2014; Vink\'o et al. 2015). In these cases it has been suggested that the observed low temperatures arise due to the reprocessing of the radiation within a thick wind (Ulmer \& Loeb 1997; Strubbe \& Quataert 2009; Lodato \& Rossi 2011; Miller 2015; Metzger \& Stone 2015; Miller et al. 2015); see, however, Piran et al. (2015) for an alternative explanation to the TDE observations which does not involve such a wind. 

In another context, Kashiyama \& Quataert (2015) considered the wind from a fallback accretion in failed explosions of compact stars as a model for rapid luminous blue transients detected by Pan-STARRS and Palomar Transient Factory. In addition, it is likely that winds can emerge from another super-Eddington accretion system -- the remnant disk in the aftermath of neutron star - neutron star or neutron star - black hole mergers (e.g., Fern\'{a}ndez \& Metzger 2013; Fern\'{a}ndez et al. 2015). Dotan, Rossi \& Shaviv (2011) and Fiacconi \& Rossi (2016) considered the the accretion-driven mass loss from a quasi-star, a proposed type of progenitor objects for supermassive black hole seeds at $z \sim 10 - 20$.        

So far, most of the wind model studies have focused on the emission property of the wind, and less is known concerning the dynamics of wind (i.e., its speed and mass loss rate). Consequently, the outflow speeds adopted in these models were very uncertain or ad hoc, ranging from very slow (e.g., nearly static in Coughlin \& Begelman 2014) to intermediate (e.g., Metzger \& Stone 2015), and to very fast (e.g., Strubbe \& Quataert 2009). 

The observed emission of a wind arises from an outer region where it is nearly optically thin so that photons diffuse out easily. Deeper inside, those photons (though not as the same as those escaping because the adiabatic cooling changes their number and energies) are trapped within the wind and dominate the internal pressure, which in turn governs the dynamics (acceleration and coasting) of the wind. Therefore, the emission properties and the dynamics are intrinsically tied together. In order to determine both correctly, a holistic approach must be taken to investigate this problem. 

We study this problem using a simple spherical model. We assume the internal pressure is dominated by the radiation. We further assume a steady state, which is justified as long as the radial expansion time is much less than the duration of the mass ejection. We characterize the wind by two free global parameters: the mass loss rate and the total specific energy. Then the radial structure of the wind can be solved as an initial value problem.  

The diffusion of photons is a loss term in energy. This impacts on both the dynamics and the emission of the wind. The inclusion of diffusion differentiates our solution from the classical adiabatic solution developed by Parker (1965; see Holzer \& Axford 1970 for a review) for the solar wind. In a reversed configuration such as the Bondi accretion problem, the photon diffusion also has a different impact on the flow, if the radiation pressure dominates over the gas pressure; see Begelman (1978). 

Our work is an extension of Meier's (1979; 1982) who studied the problem in the presence of a super-Eddington accretion disk. The radiation pressure dominated case, that we consider here, is easier to follow and it is likely more relevant for hot X-ray sources. We present numerical solutions that corroborate the analytical results, and we relate the findings to observations and apply them to a wind-dominated candidate source.

The layout of the paper is as follows. In \S 2 we describe the equations that govern the structure of the wind. In \S 3 we define four characteristic radii that shape the properties of the solutions, then we show how to analytically estimate the location of these radii, which helps us identify two regimes of solutions and delineate their asymptotic behavior. In \S 4 we present the numerical solutions. In \S 5 we apply our results to a candidate source, M101 ULX-1. We summarize the results in \S 6.  

\section{Equations}

We consider a time independent, spherical wind. Its evolution is determined by the steady-state conservation equations for mass, momentum and energy
\beq
\begin{cases}
\nabla \cdot (\rho \mathbf{v})= 0,\\
\rho (\mathbf{v} \cdot \nabla) \mathbf{v}= -\nabla P + \mathbf{f},\\
\nabla \cdot [(\frac{\rho}{2}v^2+u+P)\mathbf{v}]= \mathbf{f}\cdot\mathbf{v}-\nabla \cdot \mathbf{F},
\end{cases}
\eeq
where $\rho$ is the density, $v$ is the speed, $u$ is the internal energy density, $\mathbf{f}$ is the external gravitational force of a central point mass $M$, $\mathbf{F}$ is the radiative flux vector, and $-\nabla \cdot \mathbf{F}$ is the rate at which the radiation energy is gained by absorption and lost by re-emission. The internal energy density is 
\beq
u= u_g + u_r= 3P_g/2 + 3P_r,
\eeq
and the pressure is
\beq
P= P_g +P_r= \rho N_A k T/\mu + 4 \sigma T^4 / (3 c),
\eeq
where $N_A$ is Avogadro's number, $k$ is Boltzmann's constant, $\mu$ is the mean molecular weight,  $\sigma$ is the Stefan-Boltzmann constant, and $c$ is the speed of light.
For optically thick, radiation pressure dominated wind, which is relevant to hot, X-ray sources with high mass loss rate, the gas pressure $P_g$ is negligible, thus, we use $P= P_r$ and $u= u_r$. However, in principle, when the mass loss rate is small, the wind may become optically thin at a small radius and the gas pressure might become important in this case. Though not our focus, we briefly discuss this regime in \S \ref{sec:xsc}.

Imposing spherical symmetry, the mass conservation equation becomes
\beq     \label{eq:mass-cons}
{d \over dR} (\rho v R^2)= 0,
\eeq
where $R$ is the radius. The mass outflow rate is $\dot{M}= 4\pi R^2 \rho v$. The other two become
\beq   
{d \over dR} \left(\frac{v^2}{2}\right) + {GM \over R^2} + {1 \over \rho} {d P \over dR}= 0,
\eeq
\beq     \label{eq:energy-diff}
\frac{d}{dR} \left( \frac{v^2}{2}  + \frac{4P}{\rho} \right) + \frac{GM}{R^2} + \frac{1}{\dot{M}} \frac{d L}{dR}= 0.
\eeq
The photon diffusive luminosity $L$ is given by the diffusion approximation 
\beq   \label{eq:diff-apprx}
L= - \frac{4\pi R^2 c}{\rho \kappa_{\rm es}} \frac{d P}{d R},
\eeq
where $\kappa_{\rm es}$ is the electron scattering opacity. We assume throughout the paper that the gas is fully ionized so the opacity is dominated by electron scattering. The radiative flux and the luminosity are related by $L= 4\pi R^2 F$.

\subsection{The normalized equations}	\label{sec:normalized}

To convert the equations to a dimensionless form, we define $x \equiv R/R_g$, where $R_g= GM/c^2$ is the gravitational radius of the central object, and normalize all quantities of specific energy by $c^2$: $(v^2, P/\rho)$ $\longleftarrow$ $(v^2, P/\rho)/c^2$. Table \ref{tab:symbols} summarizes the definitions of the most of quantities used in the paper. We use the prime sign to represent the radial derivative: $()' \equiv d()/dx$. 
After these, equations (\ref{eq:mass-cons}--\ref{eq:diff-apprx}) become
\beq      \label{eq:mass}
(\rho v x^2)'= 0,
\eeq
\beq      \label{eq:mom}
\left(\frac{v^2}{2}\right)' + \frac{1}{x^2} +\frac{P'}{\rho}= 0,
\eeq
\beq     \label{eq:dif-energy}
\left(\frac{v^2}{2}\right)' + \frac{1}{x^2} +  4 \left(\frac{P}{\rho}\right)' + \frac{l'}{\dot{m}}= 0.
\eeq
\beq      \label{eq:l}
l= -x^2 P'/\rho,
\eeq
where $l \equiv L/L_{\rm Edd}$ and $\dot{m} \equiv \dot{M}/(L_{Edd}/c^2)$ are the luminosity and mass outflow rate normalized by the Eddington values, respectively. The local sound speed squared is $c_s^2= 4P/3\rho$. 

The energy equation (\ref{eq:dif-energy}) can be written in a more useful, integrated form as
\beq     \label{eq:int-energy}
\frac{v^2}{2} - \frac{1}{x} + 3 c_s^2 + \frac{l}{\dot{m}}= e,
\eeq
where the constant $e$ is the total energy per unit mass normalized by $c^2$. Consequently, $\dot{m}e$ is the total power of the wind normalized by $L_{\rm Edd}$. Plugging $P'/\rho$ from the momentum equation (\ref{eq:mom}) into equation (\ref{eq:dif-energy}) and using equation (\ref{eq:mass}), we obtain the wind equation
\beq     \label{eq:wind}
(v^2-c_s^2) \frac{d \ln v}{d \ln x}= 2 c_s^2 - \frac{1}{x} + \frac{x l'}{3\dot{m}}.
\eeq 

Note that without the flux divergence term on the right hand side, equation (\ref{eq:wind}) becomes the Parker's  equation (Parker 1965):
\beq      
\frac{xl'}{3\dot{m}} \rightarrow 0  ~~\Longrightarrow~~ (v^2-c_s^2) \frac{d \ln v}{d \ln x}= 2 c_s^2 - \frac{1}{x},   \nonumber
\eeq
albeit here the pressure is dominated by the radiation rather than by the gas; this is also the equation that governs the Bondi accretion problem (e.g., see Frank et al. 2002). In this case $x= 1/(2 c_s^2)$ is a critical point, where $v=c_s$. The existence of a critical point whose exact location is known greatly simplifies the task of finding the solution. In contrast, here in equation (\ref{eq:wind}) the existence of the flux divergence term, even when $xl'/\dot{m} \ll c_s^2$, makes it impossible to determine the position of the critical point, and clearly it is not located at $v= c_s$. Therefore, we have to use a more cumbersome numerical procedure to obtain the numerical solution (see \S\ref{sec:result}).

Combining the momentum equation (\ref{eq:mom}) with equation (\ref{eq:l}), we find
\beq     \label{eq:wind-rad}
\left(\frac{v^2}{2}\right)'= \frac{(l-1)}{x^2}.
\eeq
This equation will be helpful later in identifying the scaling of some quantities.

 \begin{table}
   \caption{Definitions of some symbols.}
    \begin{tabular}{ll}     
         \hline \hline
$M$  &  mass of the central object \\
$L_{\rm Edd}$  &  Eddington luminosity \\
$L$	&  diffusive luminosity of radiation \\
$\dot{M}$	& outflow's mass loss rate \\
$L_k$	& outflow's kinetic luminosity; $\equiv \dot{M} v^2/2$ \\
$R$	& radius \\
$v$	& outflow's speed; becoming dimensionless \\
	& (normalized by $c$) since \S \ref{sec:normalized} \\
$P/\rho$	& ratio of pressure over density; becoming \\
		& dimensionless (normalized by $c^2$) since \S \ref{sec:normalized} \\
$c_s$	& sound speed; becoming dimensionless \\
		& (normalized by $c$) since \S \ref{sec:normalized} \\
$x$	& dimensionless radius; $\equiv R/(GM/c^2)$ \\
$l$	& $\equiv L/L_{\rm Edd}$ \\
$\dot{m}$	& $\equiv \dot{M}/(L_{\rm Edd}/c^2)$ \\
$e$	& dimensionless (normalized by $c^2$) total energy per \\
	& unit mass of the outflow \\
$x_s$	& sonic radius \\
$x_t$	& photon trapping radius \\
$x_{\rm ad}$	& adiabatic radius \\
$x_{\rm sc}$	& last scattering radius (location of the scattersphere) \\
\hline
    \end{tabular}      \label{tab:symbols}
\end{table}


\section{Analytic considerations} \label{sec:anal} 

\subsection{Characteristic radii}	\label{sec:radii}

The physics of the solution  is described by  four characteristic radii. The first radius, which is the only one that exists in the radiation-free problem, is the sonic radius $x_s$. It is defined as the radius where $v=c_s$. In Parker's wind solution, the sonic point is not only a critical point, but also a dynamically pivotal point where the solution changes its scaling behavior, i.e.,  $v$ increases as $\propto x$ below $x_s$, then becomes constant after $x_s$. In our case with diffusion, the sonic point is still a pivotal point only if the wind is still adiabatic when it reaches $x_s$ (see below in \S \ref{sec:x_s}). At small radii, $v \ll c_s$, and $c_s^2$ and $1/x$ dominate equation (\ref{eq:int-energy}). As is shown by the numerical solutions in Figure \ref{fig:num-AB} these two terms are comparable wherever the radiation is coupled to the matter and provides a pressure. This happens until the last scattering radius, $x_{\rm sc}$, defined below as where the optical depth $\tau_{\rm es} (x_{\rm sc})=1$. At the sonic point $v^2= c_s^2 \approx 1/x$, therefore the wind attains its local escape speed there and it can be considered as the ``escape radius'' as well. 

The next three radii characterize the interaction of the radiation and the matter flow. The first two are determined by the optical depth: $\tau_{\rm es}(R) \equiv \int_{R}^{\infty} \rho \kappa_{\rm es} dr \simeq \rho \kappa_{\rm es} R$. The last scattering radius $x_{\rm sc}$, or the ``scattersphere'' as Meier (1982) called it, is where the scattering optical depth $\tau_{\rm es}(x_{\rm sc})= 1$, which gives $x_{\rm sc} \simeq \dot{m}/v$. The solution changes its character at  the scattersphere. The radiation field becomes very anisotropic and  the radiation pressure $P$ vanishes, even though the photons still exert a force on the matter.  In this work we consider only the massive wind regime $\dot{m} > 1$. We show later in \S \ref{sec:xsc} that this corresponds to the situation in which  $x_{\rm sc}$ is the largest of the four characteristic radii. Hence it has no effect on the dynamics of the wind. If $\dot{m} < 1$, $x_{\rm sc}$ is smaller than  other characteristic radii and, as we show in \S \ref{sec:xsc}, the solution changes drastically.

Photon diffusion becomes important, in the sense of carrying energy, at the photon trapping radius, $x_t$. At this point the luminosity of photons advected with the flow, $L_a \equiv 4(P/\rho)\dot{M}$, which is ever decreasing with $x$, equals the diffusive luminosity of photons, $L$. Below $x_t$, the radiation is trapped, the advective luminosity of photons is larger than the diffusive one ($L_a > L$), and the flow is adiabatic.

\begin{figure*}
\centerline{\includegraphics[height=14cm, angle=0]{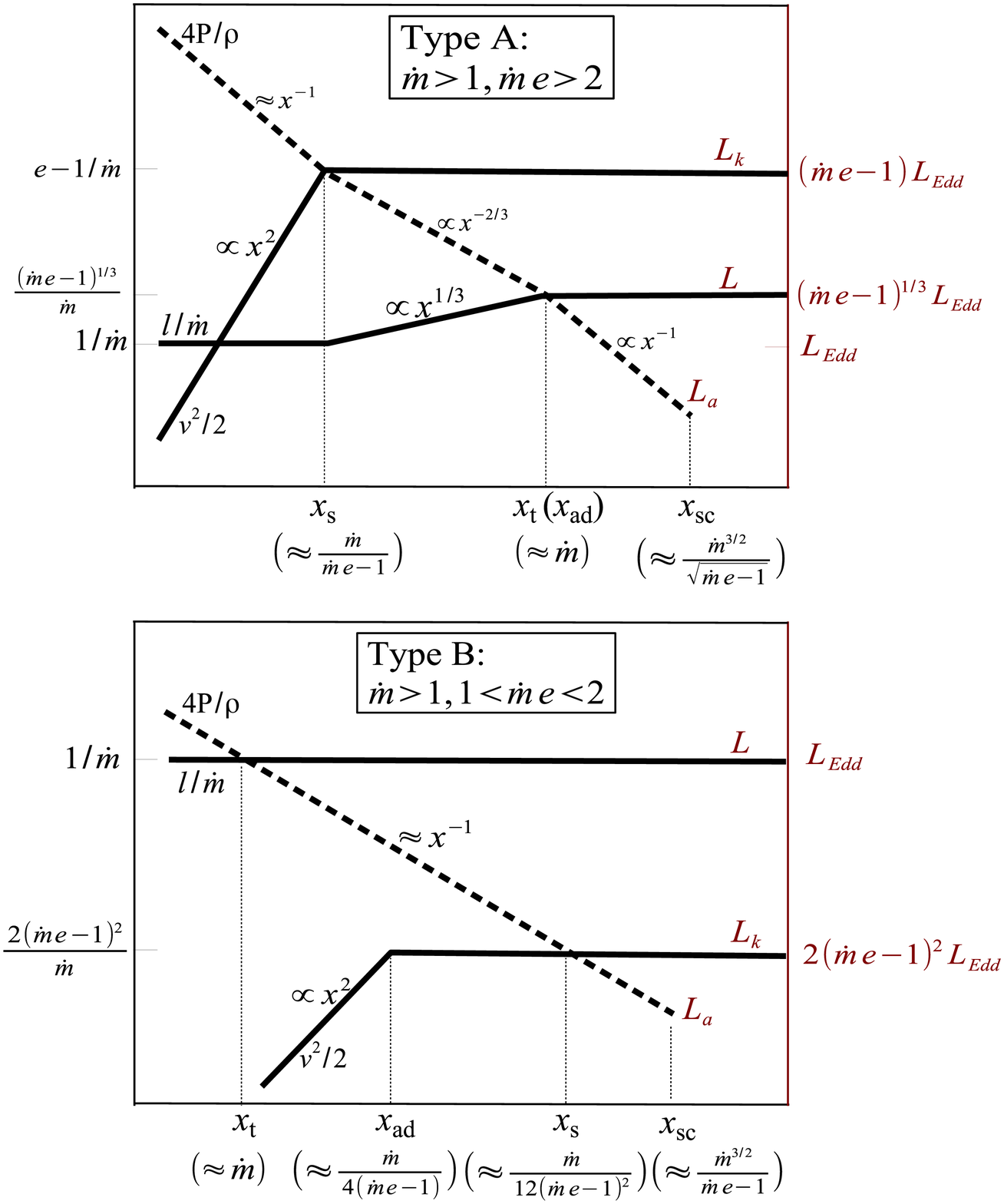}}
\caption{Schematic wind solutions. The mass loss rate (hence $x_t$) decreases from types A to B. In each panel, the $y$-axis on the left is in units of energy per unit mass normalized by $c^2$, while the $y$-axis on the right (in red color) is in absolute units of luminosity. The boundary between types A and B corresponds to $\dot{m}e= 2$. Right on this boundary, $x_s$, $x_t$ and $x_{\rm ad}$ all converge to a single radius at which $L_a= L_k=L=L_{\rm Edd}$.}   \label{fig:asymptotic}
\end{figure*}

The adiabatic radius, $x_{\rm ad}$, is defined as the radius up to which the flow is adiabatic, 
i.e., $P/\rho^{4/3}=$ constant. In the case of $x_{\rm ad} < x_s$ (so that $P/\rho \simeq 1/x$), this implies $v$ increases linearly with $x$ up to $x_{\rm ad}$. The adiabatic radius is either coincident with, or larger than, the photon trapping radius. The flow must be adiabatic up to $x_t$. However, as pointed by Meier (1982), the flow can still be adiabatic even beyond $x_t$ if the divergence of the diffusive flux $l'$ is sufficiently small. To show this, we subtract the momentum equation (\ref{eq:mom}) from the energy equation (\ref{eq:dif-energy}), obtaining
\beq     \label{eq:define-xad}
-3\left(\frac{P}{\rho}\right)'+ \frac{P}{\rho} \left(\frac{\rho'}{\rho}\right)= \frac{l'}{\dot{m}}.
\eeq
The three dashed lines plotted in Figure \ref{fig:num-AB} represent the three terms above, respectively. They show that, at small radii (e.g., $x<x_t$), the $l'/\dot{m}$ term is much smaller in absolute value than the other two on the left hand side (l.h.s), hence, $P \propto \rho^{4/3}$, i.e., the flow is adiabatic. However, as the radius increases, the terms on the l.h.s. continuously decrease while $l'/\dot{m}$ either grows or decreases slower. At some radius -- represented by $x_{\rm ad}$ -- the three terms are comparable and they remain comparable afterwards. Although the flux divergence term is small and $l$ is largely constant, this small deviation of $l$ from being exactly a constant causes the flow to deviate from $P \propto \rho^{4/3}$, i.e., from being adiabatic, and this is the essence of our type B solution (classified later; see Figures \ref{fig:asymptotic} and \ref{fig:num-AB}). 

We turn now to identify the location of the different radii and to explore the asymptotic behaviour of the solution.

 \begin{table*}              
 \caption{Asymptotic characteristics of wind solutions, complementary to Figure \ref{fig:asymptotic}.}
 \begin{tabular}{c|c|c|c|c|c} 
 \hline \hline
Types      & \multicolumn{3}{c|}{A}   & \multicolumn{2}{c}{B}  \\ \hline
\multicolumn{6}{c}{Ordering and values of radii} \\
\hline
     & \multicolumn{3}{c|}{$x_s < x_t \simeq x_{\rm ad} < x_{\rm sc}$}   & \multicolumn{2}{c}{$x_t < x_{\rm ad} < x_s < x_{\rm sc}$}   \\ \hline

$x_t$  &    \multicolumn{3}{c|}{$\dot{m}$} & \multicolumn{2}{c}{$\dot{m}$}   \\   
$x_s$  &    \multicolumn{3}{c|}{$\frac{\dot{m}}{\dot{m}e-1}$} & \multicolumn{2}{c}{$\frac{\dot{m}}{12(\dot{m}e-1)^2}$}  \\    
$x_{\rm ad}$  &    \multicolumn{3}{c|}{$\dot{m}$} & \multicolumn{2}{c}{$\frac{\dot{m}}{4 (\dot{m}e-1)}$}     \\  
$x_{\rm sc}$  &    \multicolumn{3}{c|}{$\frac{\dot{m}^{3/2}}{\sqrt{\dot{m}e-1}}$} & \multicolumn{2}{c}{$\frac{\dot{m}^{3/2}}{\dot{m}e-1}$}  \\ 
\hline
\multicolumn{6}{c}{Values and scaling of quantities} \\
\hline
	&  $x<x_s$   &    $x_s < x < x_t$    &    $x_t < x$    &   $x< x_{\rm ad}$   &  $x_{\rm ad} < x$   \\  
	\hline
$\frac{4P}{\rho}$  &  $\frac{1}{x}$  & $\propto \frac{1}{x^{2/3}}$ &  $\propto \frac{1}{x}$ & $\frac{1}{x}$ & $\frac{1}{x}$    \\  
$\frac{v^2}{2}$    & $\propto x^2$  & $e-\frac{1}{\dot{m}}$ &  $e-\frac{1}{\dot{m}}$  &  $\propto x^2$  &  $\frac{2 (\dot{m}e-1)^2}{\dot{m}}$  \\ 
$L_k / L_{\rm Edd}$  & $\propto x^2$   & $(\dot{m}e-1)$   &   $(\dot{m}e-1)$   &   $\propto x^2$   &   $2 (\dot{m}e-1)^2$   \\	
$L / L_{\rm Edd}$  &  1  &   $\propto x^{1/3}$    &   $(\dot{m}e-1)^{1/3}$    &   1   &    1    \\
\hline
\end{tabular}      \label{tab:asymptotic}
\end{table*}

\subsubsection{The photon trapping radius}	\label{sec:x_t}

The photon trapping radius is defined by the equality $L_a= L$. Approximating $d P/d R$ in equation (\ref{eq:diff-apprx}) by $-4P/R$ (this is for the case $x_t < x_s$; for $x_s < x_t$, it is $-3P/R$), one obtains $x_t \simeq \dot{m}$. The scattering optical depth at $x_t$ is $\tau_{\rm es}(x_t) \simeq 1/v$ (the speed is already normalized by $c$).  

\subsubsection{The sonic radius}		\label{sec:x_s}

For $x_s < x_t$, the flux divergence term on the r.h.s. of equation (\ref{eq:wind}) is negligible at $x \approx x_s$. This implies that the wind equation is of adiabatic form, and its solution is analogous to Parker's wind. Thus, $x_s$ is a dynamically pivotal point. Plugging $x_s= 1/(2c_s^2)$ (see discussion under equation \ref{eq:wind}) and $v=c_s$ into equation (\ref{eq:int-energy}), we obtain the location of the sonic radius $x_s \approx \dot{m}/(\dot{m}e-l) \approx \dot{m}/(\dot{m}e-1)$. The last step uses $l \approx 1$. This estimation of $x_s$ suggests that the case of $x_s < x_t$ corresponds to $\dot{m} > 1$ and $\dot{m}e>2$, i.e., a ``strong wind''.

For $x_t < x_s$, the first three terms on the l.h.s. of equation (\ref{eq:int-energy}) are all comparable around $x_s$. Thus, one has $x_s \approx (e-l/\dot{m})^{-1} = \dot{m}/[(\dot{m}e-1)-(l-1)] > \dot{m}/(\dot{m}e-1)$; the last step is obtained because $(\dot{m}e-1) \approx (l-1)$ and both are small. In this case, neither $x_t$ or $x_s$ is the dynamically pivotal point, and the real one remains to be seen. As will be shown by a more exact estimation of $x_s$ to be given below after equation (\ref{eq:rwt-energy}), this case corresponds to $\dot{m} >1$ and $1 < \dot{m}e < 2$, i.e., a ``mild wind''. The constraint on $\dot{m}$ comes from a generic requirement that $x_t > 1$.

\subsubsection{The adiabatic radius}

We turn now to estimate the adiabatic radius $x_{\rm ad}$. If $x_t < x_s$ (a mild wind, and this case corresponds to the numerical solution in Figure \ref{fig:num-AB}b), suppose that the flow remains adiabatic up to a region above $x_t$ but below $x_s$, then $P/\rho \approx 1/x$ and $v \propto x$. Therefore, $|3(P/\rho)'| \approx |(P/\rho) (\rho'/\rho)| \approx 3(P/\rho)/x$, while from equation (\ref{eq:wind-rad}) we have $l'/\dot{m} \equiv (l-1)'/\dot{m} \approx 3v^2/\dot{m}$, in which the last step uses the fact that both $(l-1)$ and $v^2$ increase as power laws of $x$ up to $x_{\rm ad}$. Thus, $x_{\rm ad}$ is a point where $(P/\rho)/x \approx v^2/\dot{m}$, or 
\beq     \label{eq:xad}
x_{\rm ad} \approx \frac{3}{4}  x_t \frac{c_s^2(x_{\rm ad})}{v^2(x_{\rm ad})}~ ~~ (\mbox{for}~x_t < x_s).
\eeq
As shown in the lower panel of Figure \ref{fig:num-AB}, in this case, $x_t < x_{\rm ad} <x_s$ and the dynamically pivotal point is not at $x_s$, but at $x_{\rm ad}$; $v$ levels off at $x_{\rm ad}$ while $P/\rho$ continues to drop as $x^{-1}$. Thus, $x_{\rm ad}^2 \approx 3x_t x_s/4$. We can estimate $x_{\rm ad}$ and $x_s$ in terms of $\dot{m}$ and $e$ in this case. At $x_{\rm ad}$, by the definition of $x_{\rm ad}$ (where the three terms in equation \ref{eq:define-xad} are comparable), $l'/\dot{m} \equiv (l-1)'/\dot{m} \approx 3 (l-1)/(\dot{m} x_{\rm ad}) \approx 3 (P/\rho)/x_{\rm ad} \approx 3/(4 x_{\rm ad}^2)$. Thus, $(l-1)/\dot{m} \approx 1/(4 x_{\rm ad})$. Rewrite equation (\ref{eq:int-energy}) to 
\beq		\label{eq:rwt-energy}
\frac{v^2}{2} - \frac{1}{x} + 4 \frac{P}{\rho} + \frac{(l-1)}{\dot{m}}= e - \frac{1}{\dot{m}}. 
\eeq
At $x_{\rm ad}$, on the l.h.s., $v^2/2$ is negligible while $4P/\rho$ cancels out with $1/x_{\rm ad}$. Therefore, we have $x_{\rm ad} \approx \dot{m}/[4 (\dot{m}e-1)]$. Now using $x_{\rm ad}^2 \approx 3 x_t x_s / 4$, we get $x_s \approx \dot{m} / [12 (\dot{m}e-1)^2]$. 

If $x_s < x_t$ (a strong wind), then at the region above $x_s$, both $P/\rho$ and $\rho$ drop with radius as power laws, thus, we still have $|3(P/\rho)'| \approx |(P/\rho) (\rho'/\rho)| \approx (P/\rho)/x$. To estimate $\l'/\dot{m}$ we cannot use equation (\ref{eq:wind-rad}) because $v$ is constant after the sonic point. Instead, we use equation (\ref{eq:l}) that defines $l$, which gives $l'/\dot{m} \approx (P/\rho)/\dot{m}$. The choice of using equation (\ref{eq:l}) assumes that $l$ must not be constant in this case (see type A in Figure \ref{fig:asymptotic}). Therefore, 
\beq   
x_{\rm ad} \approx x_t \simeq \dot{m} ~ ~~ (\mbox{for}~x_s < x_t).
\eeq
Figure \ref{fig:num-AB}a shows a numerical solution of this case.

\subsubsection{The scattersphere}     \label{sec:xsc}

The scattersphere is located at $x_{\rm sc} \simeq \dot{m}/v \simeq x_t/v$. If $\dot{m} > 1$, then $x_{\rm sc}$ lies above the dynamically pivotal point such that both $v$ and $(l-1)$ have already leveled off before reaching $x_{\rm sc}$. In this case, the scattersphere has no effect on the wind dynamics. If $x_s < x_t$, then the dynamically pivotal point is at $x_s$, thus, $x_{\rm sc} \approx \dot{m}/v(x_s) \approx \dot{m} \sqrt{x_s} \approx \dot{m}^{3/2}/ (\dot{m}e-1)^{1/2}$. If $x_t < x_s$, then the pivotal point is $x_{\rm ad}$. The speed $v$ rises as $\propto x$ until reaching $x_{\rm ad}$ and then it levels off, while $c_s^2 \approx 1/x$ is ever decreasing. Equation (\ref{eq:xad}) gives $v(x_{\rm ad}) \approx \sqrt{x_t}/x_{\rm ad}$. Hence, $x_{\rm sc} \approx \dot{m}/v(x_s) \approx x_{\rm ad} \sqrt{x_t} \approx \dot{m}^{3/2}/(\dot{m}e-1)$. 

In the case of a light wind ($\dot{m} < 1$),  $x_{\rm sc}$ is below $x_{\rm ad}$ and $x_s$. Since the solution changes its character at $x_{\rm sc}$, these two radii, $x_{\rm ad}$ and $x_s$, become meaningless. Below $x_{\rm sc}$ the equations still hold and $v \propto x$. At $x_{\rm sc}$, $v$ is still much smaller than the local escape speed.  At $x> x_{\rm sc}$ all equations other than equation (\ref{eq:wind-rad}) breaks down, as the radiation pressure vanishes. We define the gas sonic radius $x_{\rm sg}$ as the place where the wind speed $v$ equals the gas sonic speed $c_{\rm sg}= (P_g/\rho)^{1/2}$. If $x_{\rm sg} > x_{\rm sc}$, then the gas pressure is still large enough to accelerate the wind, and from there on we recover the original Parker solution. Otherwise, the radiation still exerts a force on the matter which might be enough to push it to infinity (note that we consider super-Eddington luminosities). Since we are interested in the $\dot{m} > 1$ case, we do not discuss this case further.

\begin{figure}
\centering
\includegraphics[width=8cm]{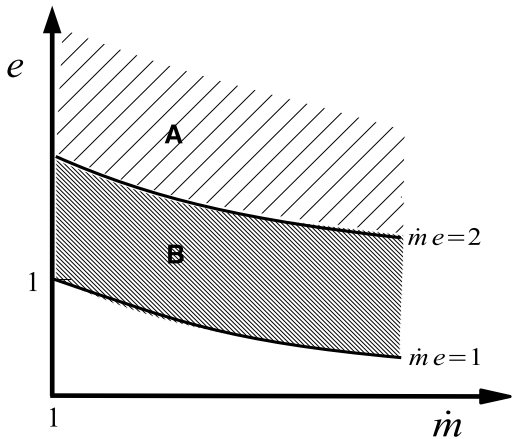}
\caption{Parameter space of the wind solutions.}   \label{fig:space}
\end{figure}

\subsection{Asymptotic behavior and the classification of the solutions}       \label{sec:asymp}

The previous analysis already delineates the asymptotic behavior of the wind solutions, which matches our numerical results, discussed below. Overall, all the solutions can be classified into two types depending on the relative locations of the sonic and trapping radii. Type A is for $x_s < x_t \approx x_{\rm ad}$, and type B is for $x_t < x_{\rm ad} < x_s$. In terms of model parameters, type A corresponds to $\dot{m} > 1$ and $\dot{m}e> 2$ and hence to a strong wind; type B corresponds to $1 < \dot{m}e < 2$ and $\dot{m} > 1$. A third regime in the parameter space, $\dot{m}<1$, corresponds to optically thin winds, in which $x_{\rm sc} < x_{\rm ad}$ (see \S \ref{sec:xsc}). 

Our types A and B correspond to Meier's (1982) ``D'' and ``C1'' solutions, respectively, for which the gas pressure $P_g$ is negligible. Meier considered two additional cases, for which $P_g$ is important -- namely, when at $x=\min(x_{\rm ad}, x_{\rm sc})$ the flow has not reached the gas sound speed $c_{\rm sg}$, yet. Those two cases describe optically thin, low temperature ($10^{4 - 5}$ K) winds, and are less relevant to X-ray sources with high mass loss rates in which we are interested here. 

With the locations of the four characteristic radii estimated in \S \ref{sec:radii}, we can obtain the scalings of the radiative luminosity $L$ and the terminal kinetic luminosity $L_k \equiv \dot{M} v^2/2$. The two types, including their values of various radii and quantities, are depicted analytically in Figure \ref{fig:asymptotic} and summarized in Table \ref{tab:asymptotic}. The ($e, \dot{m}$) parameter space of each type is shown in Figure \ref{fig:space}.

In type A solution, the wind power is dominated by the kinetic luminosity $L_k$. Both $L_k$ and $L$ are super-Eddington. In type B solution the radiative luminosity, which equals $L_{\rm Edd}$, dominates and the kinetic luminosity is sub-Eddington. 

\section{Numerical solutions}      \label{sec:result}

Here we present numerical results to accompany the analytical ones. Equations (\ref{eq:wind}) and (\ref{eq:wind-rad}) form an ordinary differential equation (ODE) set for $v^2$ and $(l-1)$, where $c_s^2$ is related to the two quantities via equation (\ref{eq:int-energy}). The two model parameters for the normalized equations are $\dot{m}$ and $e$. The mass,  $M$, is a third parameter that is needed to relate the normalized solution to a given situation in nature. 

For given $\dot{m}$ and $e$, the ODE can be numerically solved with any initial conditions ($v^2, l-1$) given at small $x$. However, there is only one solution, which corresponds to a unique set of initial conditions, that satisfies the boundary conditions of a wind, which are: at small $x$ (near the central black hole), $v^2 \ll c_s^2 \sim 1/x$, and at large $x$, $v^2 \gg c_s^2 \sim 1/x$ and the speed is finite there (the closeness of $c_s^2$ to $1/x$ will be seen later in the numerical solutions). We use the following procedure to obtain this solution.

We carry out the numerical integration in the range of $x \sim (1, 10^4)$. We first start the integration from $x=1$. Both $v^2$ and ($l-1$) increase initially. The solution is not sensitive to the initial ($l-1$) [i.e., it quickly stabilizes to some $(l-1) \ll 1$, then it smoothly increases], but it is sensitive to the initial $v^2$. A larger or smaller initial $v^2$ will send the solution to diverging upward (to infinity) or downward (to 0) at some intermediate $x$. In the regular wind solution $v^2$ should increase initially then level up. The diverging point slowly moves to larger $x$ as we approach the critical initial $v^2$. Nevertheless, as we are doing so, the part of the solution before it reaches the diverging point gives asymptotically the regular wind solution at the small $x$ region.

To obtain the regular solution at the large $x$ region, we run the integration in an opposite direction, from $x=10^4$ toward $x=1$. This time, the solution is sensitive to both the initial $v^2$ and $(l-1)$ at $x=10^4$. It also diverges at an intermediate point, and the diverging point moves leftward as the initial condition approaches a critical value. With the aid of the solution already obtained at the small $x$ region, it is easy to identify the critical initial $v^2$ and $(l-1)$ such that the large $x$ solution matches the small $x$ one.

\begin{figure*} 
\centering
\includegraphics[width=10cm, angle=0]{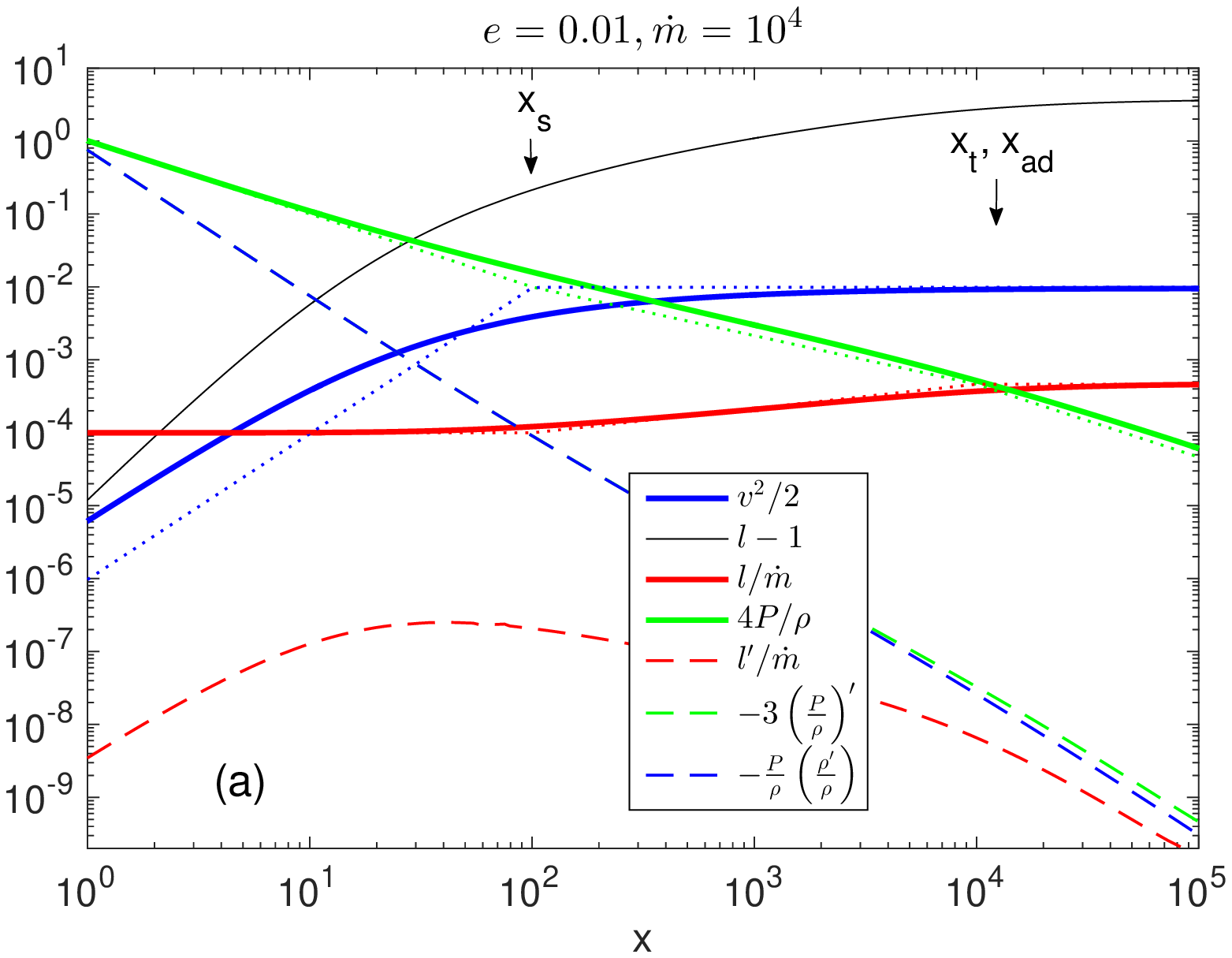}
\includegraphics[width=10cm, angle=0]{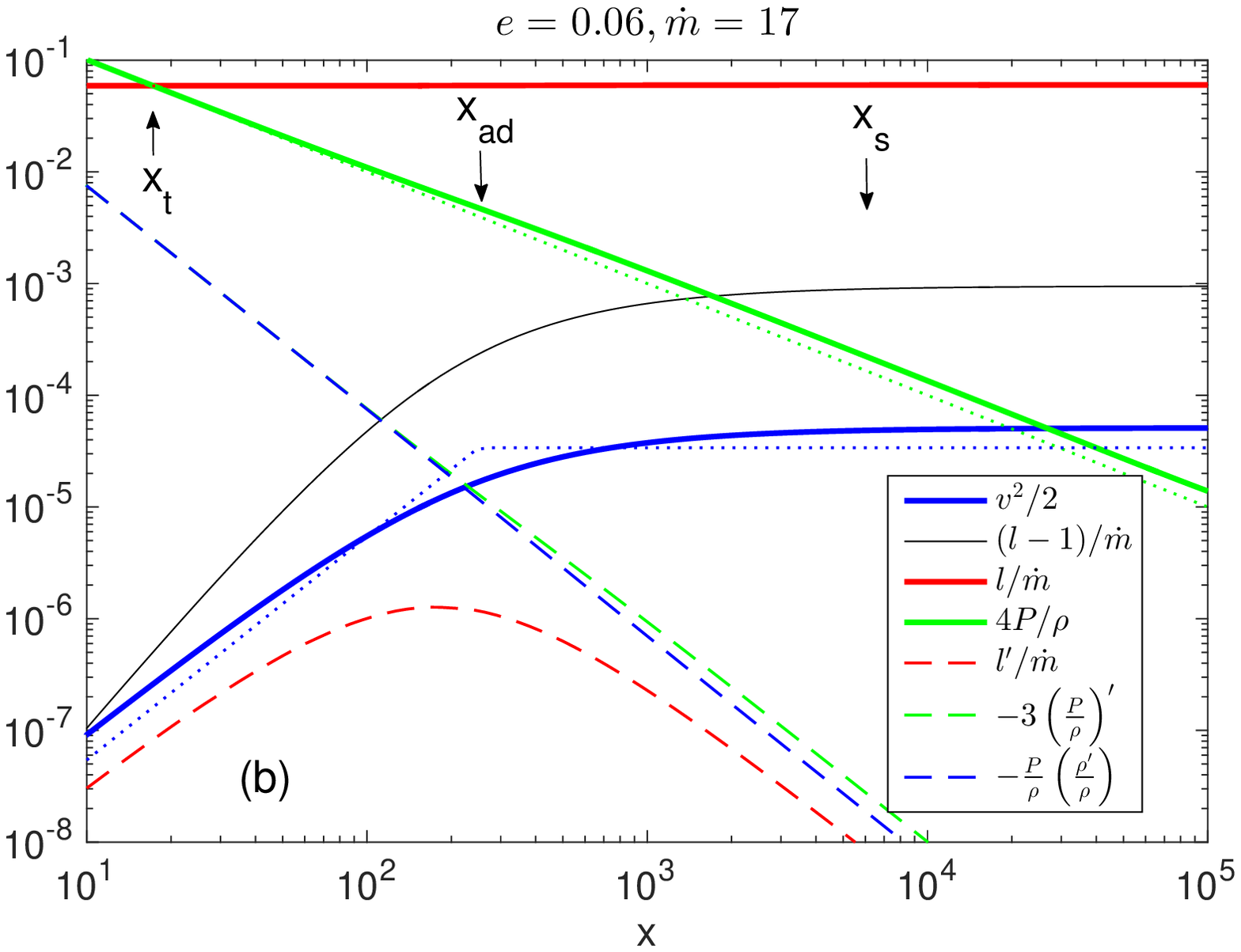}
\caption{Numerical solutions of types A (panel a) and B (panel b), respectively. The solid lines refer directly to the terms in equation (\ref{eq:int-energy}). For comparison, the dotted lines show the schematic solution presented in Table \ref{tab:asymptotic} and Figure \ref{fig:asymptotic}.  The dashed lines refer to the three terms in equation (\ref{eq:define-xad}). For panel b, the separation between $x_t$, $x_{\rm ad}$ and $x_s$ is proportional to a factor of $1/(\dot{m}e-1)$. We have chosen the value of $\dot{m}$ such that $\dot{m}e$ is very close to 1, hence, these three radii are well separated.}   \label{fig:num-AB}
\end{figure*}

This fine-tuning process is efficient such that the two diverging points, one for the small $x$ solution and one for the large $x$ solution, meet and pass well across each other. These two solutions can be combined and concatenated to give the regular solution. Figure \ref{fig:num-AB} shows examples of regular solutions and the associated quantities for two sets of ($e, \dot{m}$), respectively (corresponding to the two types of solutions discussed earlier; see \S \ref{sec:asymp}). 

\section{Application to observations}

We turn now to apply the wind solutions we have found to observations. A candidate source of the optically thick wind emission is characterized by a thermal spectrum, a high bolometric luminosity $L$ and a low color temperature $T_{\rm col}$. Following Shen et al. (2015) one can infer from $T_{\rm col}$ and $L$ the radius of the last absorption surface $R_{\rm th}$, or the ``thermalization radius'', and the mass density there $\rho(R_{\rm th})$. Assuming that the pressure is radiation dominated, one obtains $P$ at $R_{\rm th}$ from $T_{\rm col}$. Therefore, the direct and indirect observables are ($L$, $R$, $P$, $\rho$). Here and in the following, we omit the subscript of $R_{\rm th}$ and call it generally the emission radius.

On the other hand, the wind model parameters are the black hole mass, the mass loss rate and the wind's total power: ($M$, $\dot{m}$, $\dot{m}e$). The normalized emission radius $x$ is known once $M$ is known. However, so far we have only two relations connecting the observables and the model parameters, as are shown in or directly derivable from Figure \ref{fig:asymptotic} or Table \ref{tab:asymptotic}. One is for the radiative luminosity $L$, and one is for the advective luminosity $L_a$ (or $4P/\rho$). A third relation is needed. 
If $v$ or $L_k$ is available, one can easily determine the type of the solution. First, the mass loss rate is related to the speed via $\dot{M}= 4 \pi R^2 \rho v$. Then, one compares $L_k$ with $L$: if $L_k > L$, it is in type A; otherwise it is in type B. 

Alternatively the difference $4P/\rho - 1/x$ can provide this missing additional relation involving $\dot{m}$ and $e$. For type A solutions this difference is larger while for type B solutions it is very small. Therefore while it is practically impossible to measure this difference sufficiently accurate to distinguish between different type B solutions, it can be used to distinguish between type A and type B. So one can know, even without identifying the exact solution, whether radiation or kinetic energy dominates.

Furthermore, if $M$ is independently known, one can infer the type of the solution by comparing $L$ with $L_{\rm Edd}$: $L > L_{\rm Edd}$ in type A, while $L \simeq L_{\rm Edd}$ in type B.

Below, we apply this model to a source of such optically thick winds.
 
\subsection{ULX-1 in galaxy M101}

In its outbursting phase, the ultra-luminous X-ray source ULX-1 in galaxy M101 is a prototype of optically thick wind dominated sources. The binary kinetics measurement of this source constrains the black hole mass to be 20--30 $M_{\odot}$ (Liu et al. 2013). However, the combination of a high luminosity $L= 3\times10^{39}$ erg s$^{-1}$ and a low color temperature $kT=$ 0.1 keV of this source suggested that the emission radius $R$ is at least a few hundred times larger than the Schwarzschild radius of this black hole. This leaves the optically thick wind as the only possible solution for this large $R$. Deriving the wind emission radius and density from observables $L$ and $T_{\rm col}$, Shen et al. (2015) found $R \simeq 6.8\times10^9$ cm and $\rho \simeq 2\times10^{-8}$ g cm$^{-3}$. The scattering optical depth at this radius is $\tau_{\rm es}(R) \simeq 27$. 

A measurement of $v$ or $L_k$ is not available for this source, but one can infer the type of the solution. In a type A solution, the kinetic luminosity dominates over the radiative luminosity, $L_k= \dot{M}v^2/2= 2\pi R^2 \rho v^3 > L$. This requires $v > 8\times10^3$ km s$^{-1}$, and thus, $\dot{M} > 1.4\times10^{-4}~M_{\odot}$ yr$^{-1}$. This mass loss rate is significantly higher than the expected mass supply rate available from the donor star, whether it is via stellar wind capture or via Roche lobe overflow. Therefore, a type A solution is unlikely for M101 ULX-1.

A type B solution requires milder mass outflow rates. In these solutions $L \simeq L_{\rm Edd}$ so we find immediately $M \simeq 23~M_{\odot}$. This estimate is in a good agreement with the mass constraint inferred from the binary kinematics (Liu et al. 2013). Another characteristic of type B is $4P/\rho \approx 1/x$. With the values mentioned above, $1/x \equiv GM/(c^2 R) \simeq 0.5\times10^{-3}$, and $4P/\rho \simeq 1\times10^{-3}$. These two estimates are consistent with each other within a factor of 2. These two consistency checks give us confidence that a wind solution of type B is indeed a good model for this source. 

\section{Summary}

There have been growing observational hints that the emission of some luminous X-ray sources are dominated by optically thick winds. We study the dynamics and the diffusive radiation properties of such radiation-driven winds. In particular, we aim to answer those questions: What and how many types of solution are possible? What are the parameters of the wind that govern those types of solution? How and where is the terminal speed determined? What are the radiative luminosity $L$ and the kinetic luminosity $L_k$ that can be reached in such systems and, in particular, can we expect super-Eddington radiative or kinetic outflows?

We parametrize the wind by the dimensionless mass outflow rate $\dot{m}$ and the dimensionless total power $\dot{m}e$, both normalized by their Eddington values. Adopting spherical symmetry, we treat the dynamics as an initial value (at small radius) problem satisfying the boundary condition of a coasting wind (at infinity), and solve it numerically. We identify four characteristic radii, and find that all solutions can be categorized into two types, as shown in Figure \ref{fig:asymptotic}: 

\begin{itemize}
\item{A  \textit{strong} wind corresponds to $\dot{m} > 1$ and $\dot{m} e > 2$. This type of solution recovers Parker's classical, adiabatic solar wind solution. Due to the high $\dot{m}$, photon diffusion is unimportant in shaping the dynamics because the photon trapping radius $x_t$ is located outside the sonic radius $x_s$. It is the latter that marks the end of the wind's acceleration. The kinetic and radiative luminosities are both super-Eddington, i.e., $L_k > L > L_{\rm Edd}$. In this regime $L_k \approx  
\dot{m} e L_{\rm Edd}$ and $L \approx (L_k/L_{\rm Edd})^{1/3} L_{\rm Edd} \approx  (\dot{m} e)^{1/3} L_{\rm Edd}$. Thus $L$ can be larger than $L_{\rm Edd}$ but it increases slowly with the wind's total power. } 

\item{A \textit{mild} wind corresponds to a lower $\dot{m}$ (but still $> 1$) and $1 < \dot{m}e < 2$. The mass outflow rate exceeds Eddington, but the total outflow power is only slightly above Eddington. Here, photon diffusion becomes important, as the photon trapping radius $x_t$ recedes to inside $x_s$. The acceleration ceases at the adiabatic radius $x_{\rm ad}$, the radius that marks the end of wind being adiabatic and is located between $x_t$ and $x_s$. The radiative luminosity is Eddington while the kinetic one is sub-Eddington, $L \approx L_{\rm Edd} > L_k$.}

\end{itemize}

Observationally, the most probable candidate sources of such optically thick winds are those characterized by a high luminosity ($\gtrsim 10^{39}$ erg s$^{-1}$), a thermal spectrum and a low color temperature ($\lesssim 1$ keV).  Once those characteristics being satisfied, the type of solution can be identified using the relative strengths of $L$, $L_k$ and $L_{\rm Edd}$ (or $M$) as diagnostics. For instance, in a \textit{strong} wind $L_k > L > L_{\rm Edd}$. A \textit{mild} wind must have $L \approx L_{\rm Edd}$. If it is identified that a solution is in the mild regime, we can use this to estimate M, the mass of the black hole, from the radiative luminosity. This is the case we found for ULX-1 in galaxy M101.

\section*{Acknowledgments}

We thank Rodolfo Barniol Duran for helpful discussion and comments and for carefully reading the manuscript. This research was partially supported by the I-CORE Program (1829/12). TP was partially supported by an ISF-CNSF grant, and an ISA grant. EN was was partially supported by an ERC starting grant (GRB/SN), ISF grant (1277/13) and an ISA grant.







\bsp	
\label{lastpage}
\end{document}